\documentclass[fleqn,usenatbib]{mnras}

\usepackage{newtxtext,newtxmath,amsmath}
\usepackage{float}
\usepackage{amsmath,bm}
\usepackage{xcolor}

\usepackage[T1]{fontenc}
\usepackage{ae,aecompl}
\usepackage{nccmath}
\usepackage{soul}

\usepackage{graphicx}	
\usepackage{amsmath}	
\usepackage{amssymb}	

\newcommand{\nn}{\nonumber}
\newcommand{\LCDM}{\Lambda \rm CDM}
\newcommand{\fNL}{f_{\rm NL}}

\newcommand{\eqrb}{\begin{eqnarray}}
\newcommand{\eqre}{\end{eqnarray}}

\makeatletter
\newcommand*{\rom}[1]{\expandafter\@slowromancap\romannumeral #1@}
\makeatother
\graphicspath{{./figs/}}

\title[Interacting Dark Energy with Non-Gaussian Initial Conditions]{Cosmic Degeneracies III: N-body Simulations of Interacting Dark Energy with Non-Gaussian Initial Conditions}

\author[M. Hashim et al.]{
M. Hashim,$^{1,2,8}$\thanks{E-mail: mahmoudyousif.hashim@unibo.it}
C. Giocoli,$^{1,2,3}$
M. Baldi$^{1,2,3}$
D. Bertacca,$^{4,5,6,7}$
R. Maartens$^{8, 9}$ 
\\
$^{1}$Dipartimento di Fisica e Astronomia, Alma Mater Studiorum Universit\`a di Bologna, viale Berti Pichat, 6/2, I-40127 Bologna, Italy;\\
$^{2}$INAF - Osservatorio Astronomico di Bologna, via Ranzani 1, I-40127 Bologna, Italy;\\
$^{3}$INFN - Sezione di Bologna, viale Berti Pichat 6/2, I-40127 Bologna, Italy;\\
$^4$Dipartimento di Fisica e Astronomia ``G. Galilei'', Universit\`a degli Studi di Padova, via Marzolo 8,  I-35131 Padova, Italy;\\
$^5$ INFN, Sezione di Padova, via Marzolo 8, I-35131 Padova, Italy;\\
$^6$ Argelander-Institut f\"ur Astronomie, Auf dem Hugel 71, D-53121 Bonn, Germany;\\
$^7$ Departament de F\'isica Qu\`antica i Astrofis\'ica \& Institut de Ci\`encies del Cosmos, Universitat de Barcelona,\\ Mart\'i i Franqu\`es 1,  08028 Barcelona, Spain;\\
$^{8}$Department of Physics \& Astronomy, University of the Western Cape, Cape Town 7535, South Africa;\\
$^{9}$Institute of Cosmology \& Gravitation, University of Portsmouth, Portsmouth PO1 3FX, UK.
}

\date{Accepted 2018 September 5. Received 2018 August 22; in original form 2018 June 12.}

\pubyear{2018}

\begin{document}
\label{firstpage}
\pagerange{\pageref{firstpage}--\pageref{lastpage}}
\maketitle

\begin{abstract}
We perform for the first time N-body simulations of Interacting Dark Energy assuming non-Gaussian initial conditions, with the aim of investigating possible degeneracies of these two theoretically independent phenomena in different observational probes. We focus on the large-scale matter distribution, as well as on the statistical and structural properties of collapsed halos and cosmic voids. On very large scales, we show that it is possible to choose the Interaction and non-Gaussian parameters such that their effects on the halo power spectrum cancel, and the power spectrum is indistinguishable from a  $\LCDM$ model. On small scales, measurements of the non-linear matter power spectrum, halo-matter bias, halo and subhalo mass function and cosmic void number function validate the degeneracy determined on large scales. However, the internal structural properties of halos and cosmic voids, namely halo concentration-mass relation and void density profile, are very different from those measured in the $\LCDM$ model, thereby breaking the degeneracy. In practice, the values of $\fNL$ required to cancel the effect of interaction are already ruled by observations. Our results show in principle that the  combination of large- and small-scale probes is needed to constrain Interacting Dark Energy and Primordial non-Gaussianity separately.

\end{abstract}



\section{Introduction}
According to the most recent measurements of cosmic microwave background (CMB) anisotropies 
performed by the Planck satellite mission \citep{Ade:2015xua}, the standard $\Lambda $CDM cosmological model is still extremely successful in reproducing different observational datasets. This in turn favours the more economic cosmological constant $\Lambda $ as an explanation of the late-time cosmic acceleration over alternative and more complex Dark Energy (DE) or Modified Gravity (MG) models. Nonetheless, theoretical problems in understanding the energy scale and the time evolution of $\Lambda $ \citep[known as the {\em fine-tuning} and {\em coincidence} problems, respectively, see e.g.][]{Weinberg:1988cp,Padilla:2015aaa} as well as recent observational tensions between CMB cosmological constraints and those inferred from independent probes in the local Universe \citep[see e.g.][]{Heymans:2013fya,Hildebrandt:2016iqg,Simpson:2015yfa,Vikhlinin:2008ym,Ade:2015fva} motivate the investigation of such alternative and more complex scenarios. 

In particular, various possible realisations of Interacting Dark Energy models \citep[hereafter IDE, see e.g.][]{Wetterich:1994bg,Amendola:1999er,Pettorino:2008ez,Amendola:2007yx,Baldi:2010vv,Baldi:2012kt,Pourtsidou:2013nha} based on a direct energy-momentum exchange between a DE scalar field and the CDM particle sector, have attracted significant interest and for small values of the interaction strength appear still consistent with current CMB constraints \citep[][]{Salvatelli:2013wra,Costa:2013sva, Salvatelli:2014zta,Ade:2015rim}.

Similarly, measurements of higher-order statistics of the CMB anisotropies are consistent with a nearly Gaussian distribution of the primordial curvature perturbations by providing very tight constraints on the Primordial Non-Gaussianity (hereafter PNG) parameters: $f^{\rm loc}_{\rm NL} = 0.5 \pm 5.0$ and $f^{\rm eq}_{\rm NL} = -4.0 \pm 43.0$ for the local and equilateral configurations, respectively \citep{Ade:2015ava}. As some level of non-Gaussianity in the primordial density distribution is a common and clean prediction of basically all models of inflation \citep[see e.g.][]{Maldacena:2002vr} -- i.e. the hypothetical mechanism driving the early exponential expansion of the Universe -- measurements of PNG are considered as a smoking gun to discriminate between various inflationary models \citep{Bartolo:2004if,Giannantonio:2013uqa}. 

Despite the tight constraints on the PNG amplitude from Planck data, Large-Scale Structure (LSS) observations in the late universe coming from the next generation of wide-field galaxy redshift surveys could outperform these constraints. More specifically, recent measurements of galaxy clustering and of the integrated Sachs-Wolfe (ISW) effect already provide constraints of $\sigma (f^{\rm loc}_{\rm NL}) \sim 30$ \citep{Ross:2012sx, Giannantonio:2013uqa,Leistedt:2014zqa}, while future redshift galaxy surveys like {\small Euclid} \citep[][]{Laureijs:2011gra} and SKA \citep[][]{Maartens:2015mra,Camera:2014bwa} are forecast to outperform the CMB in constraining PNG, especially via the multi-tracer method \citep{Alonso:2015sfa,Fonseca:2015laa}. 

This is possible due to the various observational signatures that PNG imprints on LSS at late times, namely on the abundance of massive objects (which can be either enhanced or suppressed for positive and negative values of the PNG amplitude, respectively), on the bias between galaxies and the underlying matter distribution (that becomes scale-dependent on large scales in the presence of some PNG) and on the 3-point correlation function of galaxies that encodes the shape of PNG \citep[for more details, see e.g.][]{Desjacques:2010jw,Liguori:2010hx,Desjacques:2016bnm}.

Recent studies on the effects of IDE models on structure formation \citep[][]{Baldi:2008ay,Moresco:2013nfa,Hashim:2014rda, Duniya:2015nva, Cui:2012is} have shown that similar features may arise also in these models. In particular, \citet{Hashim:2014rda} showed that the observational signatures of IDE and PNG on the large-scale galaxy power spectrum can mimic each other.
This is due to the fact that some models of IDE introduce a $k^{-2}$ scale dependence in the matter density contrast on very large scales, mimicking PNG scale-dependent halo bias. Also on non-linear scales, numerical simulations of IDE \citep[see e.g.][]{Baldi:2010td,Baldi:2011qi,Baldi:2011th,Cui:2012is} and of PNG \citep[][]{Grossi:2007ry,Pillepich:2008ka,Wagner:2010me,LoVerde:2011iz} scenarios showed that IDE enhances the abundance of massive halos in a similar way to  PNG with a positive amplitude. 

This degenerate behaviour between PNG and IDE indicates that separate observational constraints on the PNG amplitude and the IDE interaction rate could be misinterpreted or possibly that their joint effects could become indistinguishable from the standard $\Lambda $CDM reference model. This represents the main motivation for the present work, where we will present for the first time a joint numerical analysis of non-linear structures forming from PNG initial conditions through an IDE cosmological evolution. Our main goal is to test whether such degeneracy holds for all observables at all scales and if not to identify specific statistics that clearly disentangle the two phenomena. 
To this end we will consider -- as a proof of concept -- very large values of the PNG amplitude $f_{\rm NL}\approx {\mathcal O}(100)$ which are already ruled out by CMB observations for the simple case of scale-independent non-Gaussianity. This allows us to obtain larger effects on structure formation and to identify more clearly the degeneracy with Dark Energy interactions. Although not directly applicable to realistic PNG scenarios for the case of a scale-independent $f_{\rm NL}$, our results will provide a guideline for scale-dependent PNG models \citep[see e.g.][]{Sefusatti:2009xu,Oppizzi:2017nfy} having $f_{\rm NL} \approx {\mathcal O}(100)$ or larger at the scales relevant for non-linear structure formation while remaining consistent with CMB constraints at the Planck pivot scale.

This paper is organized as follows: in Sec.~\ref{Sec-I} we introduce IDE and PNG extensions to the standard $\LCDM$ scenario. In Sec.~\ref{Sec-II} we use the linear halo power spectrum as an observational probe to test the IDE--PNG degeneracy on large scales. In Sec.~\ref{Sec-III} we test the IDE--PNG degeneracy on non-linear scales by running a set of N-Body simulations for all models under consideration. In Sec.~\ref{Sec-IV} we present all results for the non-linear matter power spectrum, halo-matter bias, halo mass function, subhalo mass function, halo concentration-mass relation, void number density and void density profiles. Finally, conclusions are summarised in Sec.~\ref{Sec-V}. 

\section{Non-Standard Cosmological Models}\label{Sec-I}

In this section, we present the two non-standard extensions to the fiducial $\LCDM$ model that we  will consider in this work. The first extension is based on the assumption of a non-Gravitational interaction between CDM particles and a dynamical DE scalar field. The other extension relays on a non-Gaussian distribution of the primordial density field as generically predicted by inflationary models.  
 
\subsection{Interacting Dark Energy}\label{Sec-I-I}

Various models of IDE have been  proposed in the literature over the past two decades \citep[see e.g.][]{Amendola:1999er, Amendola:2003wa, Koyama:2009gd, Baldi:2010vv, Clemson:2011an}. In this paper we consider the most widely studied example of such models based on a quintessence dynamical scalar field $\phi$ playing the role of the DE, subject to a self-interaction potential $V(\phi)$ and to a direct interaction with the CDM fluid via energy-momentum exchange \citep{Bertolami:1999dp, Amendola:1999er}. The background evolution of such cosmological scenarios is governed by the Klein-Gordon equation for the scalar field:
\begin{ceqn}
\begin{align}
	\ddot{\phi} + 3 H \dot{\phi} + \frac{dV}{d\phi} = \sqrt{\frac{2}{3}} \kappa \beta \rho_c, \label{kg-Eq}
\end{align}
\end{ceqn}	
and by the continuity equations of the different components that contribute to the total energy density of the universe:   
\begin{eqnarray}
&&\dot{\rho }_{c} + 3 H \rho_c = - \sqrt{\frac{2}{3}} \kappa \beta \rho_c \dot{\phi}, \label{cdm-Eq}\\
&&\dot{\rho }_{b} + 3 H \rho_b = 0, \\
&&\dot{\rho }_{r} + 4 H \rho_r = 0,
\end{eqnarray}
as well as by the Friedmann constraint 
\begin{ceqn}
\begin{align}
3H^2 = \kappa^2 \left(\rho_{\phi} + \rho_c + \rho_b + \rho_r \right),
\end{align}
\label{Hubble_fun-Eq}
\end{ceqn}
where $\rho_c,\rho_b$ and $\rho_r$ are the energy density of CDM, baryons and radiation, respectively. An overdot represents a derivative with respect to the cosmological time $t$. The Hubble function is defined as $H \equiv \dot{a}/a$ where $a$ is the scale factor and $\kappa^2 = 8 \pi G$. The parameter $\rho_{\phi}$ represents the energy density of the DE fluid defined as $\rho_{\phi} =  \dot{\phi}^2/2 + V(\phi)$. The right-hand side source terms in Eqs. \eqref{kg-Eq} and \eqref{cdm-Eq} represent the interaction parameter between CDM particles and DE that is proportional to the CDM energy density $\rho_c$ through the dimensionless constant $\beta$ that sets the strength of the coupling. The sign of the $\beta \dot{\phi }$ term determines the direction of the energy-momentum exchange between the two interacting components. In order to fulfil Bianchi identities and not violate total energy-momentum conservation, the source terms in Eqs. \eqref{kg-Eq} and  \eqref{cdm-Eq} should be equal and have opposite sign. 

By integrating the CDM conservation 
equation \eqref{cdm-Eq} one gets the time evolution of the CDM density as:
\begin{ceqn}
\begin{align}
\frac{\rho_c}{\rho_{c0}} = a^{-3} \exp \left(- \sqrt{\frac{2}{3}}\, \kappa \beta \phi \right),
\end{align}
\end{ceqn}
which shows a basic property of IDE models: matter density is not separately conserved as the energy exchange results in  a time-dependent CDM particle mass. In this work, we consider the exponential form for the self-interaction potential  \citep{Wetterich:1987fm,Lucchin:1984yf}, 
\begin{ceqn}
\begin{align}
V(\phi) = A \exp\left(- \sqrt{\frac{2}{3}}\, \kappa \lambda \phi\right),
\end{align}
\end{ceqn}
where $A$ and $\lambda$ are constants. 

In the Newtonian gauge, the perturbed metric (assuming flatness and vanishing anisotropic stress) is given by
\begin{ceqn}
\begin{align}
ds^2 = \left[ -(1 + 2\Phi) dt^2 + (1 - 2\Phi) a^2  d\pmb{x}^2 \right], 
\end{align}
\end{ceqn}
where $\Phi$ is the gravitational potential. The Poisson equation is\footnote{We also ignore baryons for simplicity.}  \citep{Hashim:2014rda}:
 \begin{ceqn}
 \begin{align}\label{Eq-PossN}
 \nabla^2 \Phi = \frac{\kappa^2}{2} \left[ \rho_c \Delta_c + \rho_{\phi} \Delta_{\phi} - \sqrt{\frac{2}{3}} \kappa \beta \dot{\phi} \frac{\rho_c}{(\rho_c + \rho_{\phi})}(v_{\phi} - v_c) \right],
 \end{align}
 \end{ceqn}
where $\Delta_{c,\phi}$ are the comoving density contrasts and $v_{c,\phi}$ are the velocity potentials, defined by
$\bm{v}_{c,\phi}=\bm{\nabla}v_{c,\phi}$, so that $\theta_{c,\phi}=-k^2v_{c,\phi}$, where $\theta$ is the velocity divergence.
The velocity potentials include a $k^{-2}$ scale-dependence due to the potential $\Phi$ in the Euler equation -- see Eq. \eqref{10} below. Therefore the coupling term in the Poisson equation \eqref{Eq-PossN} introduces a $k^{-2}$ scale-dependence to the matter growth factor on large scales. Since $\Delta_{c,\phi}$ are gauge-invariant, the resulting $k^{-2}$ signal is an explicit coupling effect and not a false gauge effect.

The perturbed conservation equations are then given by \citep{Hashim:2014rda}
\begin{eqnarray}
	&& \dot{v}_i + H v_i + \frac{c^2_{si}}{(1 + w_i)} \Delta_i + \Phi = \frac{1}{(1 + w_i) \rho_i} \Big[Q_i \left(v -  v_i \right)  \rho_c + f_i \Big], \label{10}\\
	&& \dot{\Delta}_i - 3 w_i H\Delta_i - k^2 (1 + w_i) v_i - \frac{9}{2}  H^2 (1 + w_i) (1 + w_t) ( v_i - v) = \frac{\mathcal{Q}_i^\Delta}{H},\nonumber \\
\label{CDM_pertrub-Eq}
\end{eqnarray}
where $Q_\phi =   \sqrt{2/3} \kappa \beta \rho_c \dot{\phi}=-Q_c$ and $i$ indicates CDM and scalar field $\phi$ respectively, $c_{si}$ is the sound-speed of the $i$-th species (which is vanishing for CDM while for DE perturbations $c_{s\phi} = 1$), $w_t$ is the total equation of state, $v = 1/(1 + w_t)\sum_i (1 + w_i) \Omega_i v_i$ is the total peculiar velocity potential and $f_i$ is the momentum transfer potential given by \citep{Koyama:2009gd} 
\begin{ceqn}
\begin{align}
f_i = Q_i (v_{\phi} - v)\,.
\end{align}
\end{ceqn}
The source term on the right hand side of Eq. \eqref{CDM_pertrub-Eq} is given by 
\begin{eqnarray}
	\mathcal{Q}_i^\Delta &=& \frac{Q_i }{\rho_i}\left[ \frac{\dot{Q}_i}{Q_i} - \frac{\dot{\rho}_i}{\rho_i} \right] v_i
								 - \frac{Q_i }{\rho_i} \left[ 3 + \frac{Q_i}{(1 + w_i) \rho_i H }\right] \left(v -  v_i \right)\nn \\
								&& - \frac{1}{\rho_i} \left[ 3 + \frac{Q_i }{(1 + w_i) \rho_i H}\right] f_i + \frac{Q_i}{\rho_i} \left[3 (1 + w_i) 
								+ \frac{Q_i}{\rho_i H}\right] v_i \nn \\
								&& + \frac{1}{\rho_i H}\,\delta Q_i   - \frac{Q_i}{\rho_i H} \left[ \frac{c^2_{si}}{(1 + w_i)} + 1\right] \Delta_i 
								+ 2 \frac{Q_i}{\rho_i H}\, \Phi.
\end{eqnarray}
These equations fully specify the evolution of the linear gauge-invariant perturbations of the coupled system, we refer the interested reader to \citet{Hashim:2014rda} for a more complete derivation of these equations.

As we will be interested in the evolution of the system at small scales and beyond the linear regime (see Sec. \ref{Sec-NBody} for details), we also recall \citep[see e.g.][]{Amendola:2003wa} that in the Newtonian limit, used for the N-Body implementations, the evolution equation for CDM density perturbations, Eqs. \eqref{10} and \eqref{CDM_pertrub-Eq} imply:
\begin{ceqn}
\begin{align}\label{cdm_pertrub-Eq}
\ddot{\delta}_c + 2H \left(1 -\beta \frac{\dot{\phi}}{H\sqrt{6}} \right)\dot{\delta}_c - \frac{\kappa^2}{2} \rho_c  \left(1 + \frac{4}{3} \beta^2\right)\delta_c = 0, 
\end{align}
\end{ceqn}
since comoving and Newtonian density contrasts are equal, i.e. $\Delta_c \approx \delta_c$, and we  ignore derivatives of scalar field perturbations. The coupling terms in Eq. \eqref{cdm_pertrub-Eq} are: $\beta \dot{\phi}$, which represents an extra friction arising as a consequence of momentum conservation, and  $4\beta^2/3$, which is responsible for the fifth force acting on CDM perturbations. 
  
\subsection{Primordial Non-Gaussianity}

Local type non-Gaussianity in the primordial curvature perturbations, that maximizes the bispectrum in the squeezed shape, is parametrized  by 
\begin{ceqn}
\begin{align}\label{localPNG-Eq}
\Phi = \Phi_G + f^{\rm loc}_{\rm NL}\left(\Phi^2_G - \langle \Phi^2_G \rangle \right),
\end{align}
\end{ceqn} 
where $\Phi_G$ is the Gaussian gravitational field and $f^{\rm loc}_{\rm NL}$ is the PNG parameter. Single-field inflation models predict a very small value of $f^{\rm loc}_{\rm NL}$ \citep{Maldacena:2002vr}, but multifield models can generate large non-Gaussianity in squeezed configurations \citep{Moroi:2001ct,Lyth:2001nq}. 

On large scales, PNG enhances the large peaks of matter perturbations \citep{Matarrese:2008nc,LoVerde:2007ri,Matarrese:2000iz}. This introduces a scale-dependent signal in the bias between the virial collapsed objects at high peaks and the underlying traced matter. By measuring the cross halo-matter power spectrum $P_{mh}$ in N-Body simulations with local-type non-Gaussian initial conditions, many authors have confirmed that the large-scale bias is scale-dependent \citep[see e.g.][]{Dalal:2007cu,Pillepich:2008ka}:
\begin{ceqn}
\begin{align}\label{hmpowspec-Eq}
P_{hm}(k,z) = \left[b_G(z) + \Delta b(k,z)\right] P_{mm}(k,z),
\end{align}
\end{ceqn} 
where $P_{mm}$ is the matter auto-power spectrum, $b_G$ is the Gaussian bias and
\begin{ceqn}
\begin{align}
\Delta b(k,z) = 3 f_{\rm NL} \big[b_G(z) - 1\big] \frac{\delta_{\rm crit}\,\Omega_m}{D_c(z)T(k)}\,\frac{H_0^2}{k^2},
\end{align}
\end{ceqn}
with $\delta_{\rm crit}$ being the critical overdensity for halo collapse, $T(k)$ the transfer function and $D_c$ the linear dark matter growth factor which is normalised to $a$ in the matter dominated era. On very large scales, $T \rightarrow 1$ and so $\Delta b \propto f_{\rm NL} k^{-2}$.  Since we only consider local type PNG in the current analysis, for simplicity we drop the {loc} superscript from our notation.

Since IDE introduces a scale dependence in the matter growth factor and non-negligible DE perturbations in the Poisson equation \eqref{Eq-PossN}, the scale-dependent PNG bias for IDE models becomes 
\begin{ceqn}
\begin{align}\label{bias-Eq}
\Delta b(k,z) = 3 f_{\rm NL} \big[b_G(z) - 1\big] \frac{\delta_{\rm crit}\,\Omega_m}{D_c(k,z)\big[1+\mu(k,z)\big]T(k)}\,\frac{H_0^2}{k^2},
\end{align}
\end{ceqn} 
where the effect of IDE appears in the scale dependence of $D_c$ and in the factor 
\begin{ceqn}
\begin{align}
\mu = \frac{\dot{\rho}_{\phi}}{\dot{\rho}_c}  \left[1 - \left(\frac{\dot{\rho}_{\phi}}{\dot{\rho}_c}\right) \right]^{-1} \frac{1}{D_{\phi}},
\end{align}
\end{ceqn}
where $D_{\phi} \equiv \Delta_{\phi}/\Delta_{\phi}(z = \infty)$ is the DE growth factor and $\mu$ depends on the coupling parameter $\beta$ though the background equations \eqref{kg-Eq} and \eqref{cdm-Eq}. We can notice that on very large scales, $D_c(k,z)$ behaves as  $\sim k^{-2}$. 

\section{Linear Halo Power Spectrum}\label{Sec-II}

\begin{figure*}
\centering
\includegraphics[width=\linewidth]{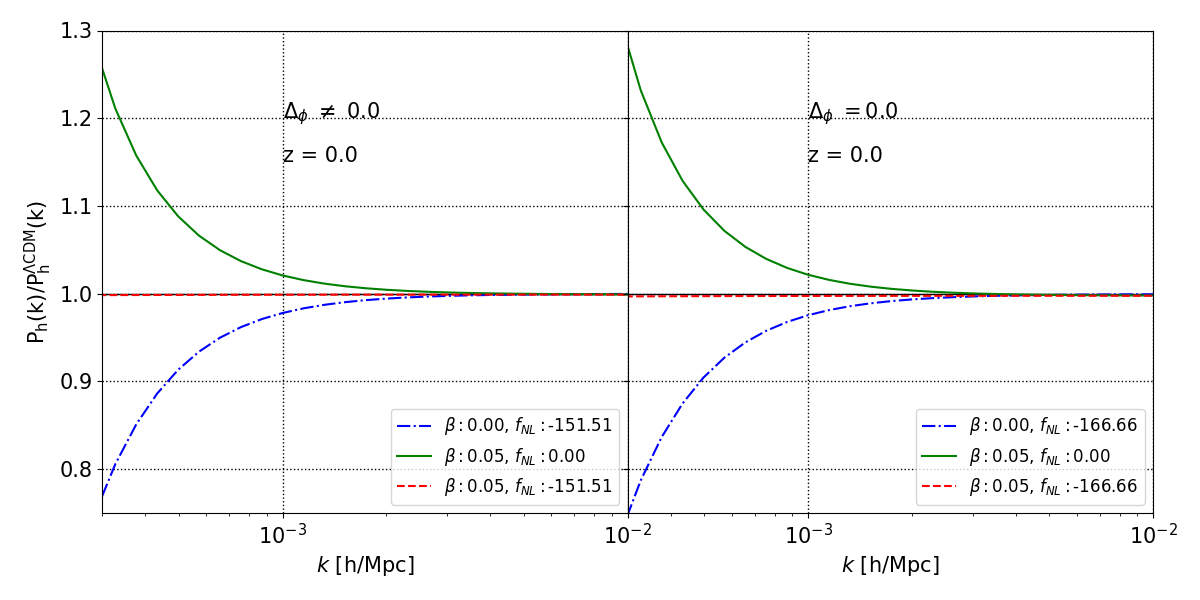}
\caption{The ratio of the linear halo power spectrum to the standard $\LCDM$ case for the models (\rom{1}--\rom{6}) given in Table \ref{Table-II} at $z = 0$, assuming perturbed DE (left panel) and non-perturbed DE (right panel).}
\label{Fig-I}
\end{figure*}

\begin{table}
\centering
\begin{tabular}{cc}
\hline
Parameter & Value\\
\hline

$h$ & 0.703 \\$\Omega_b$ & 0.0451\\
$\Omega_m$ & 0.2711\\
$\Omega_{\rm DE}$ & 0.729\\
\hline
$A_s$ & 2.42 $\times 10^{-9}$ \\
$n_s$ & 0.966\\
\hline
\end{tabular}
\caption{The cosmological parameters used in this paper, consistent with the WMAP7 CMB data best fit \citep{2011ApJS..192...18K}.}
\label{Table-I}
\end{table}

In this section, we will illustrate the degeneracy between IDE and PNG  by computing the halo power spectrum on linear scales for both models and for their combination.

The halo power spectrum is given in general by
\begin{ceqn}
\begin{align}\label{Eq-Pg}
P_{h}(k,z) = \left[b_G(z) + \Delta b(k,z)\right]^2 P_{m}(k,z).
\end{align}
\end{ceqn}
In order to compute this we first numerically solve Eqs. \eqref{10} and \eqref{CDM_pertrub-Eq} for the growth factors $D_{i}$, and then calculate the matter power spectrum $P_{m}(k,z)$ using  \citep[][]{Ade:2015lrj}:
\begin{ceqn}
\begin{align}\label{Eq-Pm}
P_m(k,z)  = A_s^2 \left(\frac{k}{k_p}\right)^{n_s} T^2(k) \left[\frac{D_c(k,z)}{D_c(k,0)}\right]^2,
\end{align}
\end{ceqn}  
where $n_s$ is the spectral index, $A_s$ is the spectral amplitude and $k_p$ is the pivot scale. We use $\rm CAMB$ \citep{Lewis:1999bs} to compute the transfer function $T(k)$. We then apply the bias relation, Eq. \eqref{bias-Eq}, to the matter power spectrum as given in Eq. \eqref{Eq-Pg}\footnote{For the Gaussian bias, we use the ansatz $b_G = \sqrt{1 + z}$.}. We adopt the cosmological parameters given in Table \ref{Table-I}.

In computing the growth rate of CDM density perturbations we consider both the case where large-scale perturbations in the DE scalar field are properly taken into account ($\Delta_{\phi} \neq 0$) and the the case where such perturbations are artificially set to zero ($\Delta_{\phi} = 0$). The latter case, while being not fully consistent, allows us to match the approximations adopted in the numerical treatment that we will discuss below and to obtain a more direct correspondence between the PNG and IDE parameters that are expected to provide a strong degeneracy in the non-linear regime under such approximations. In Fig. \ref{Fig-I}, we show the ratio of the linear halo power spectrum to the fiducial $\LCDM$ model at $z =0$ for the cases given in Table \ref{Table-II}.
\begin{table}
\centering
\begin{tabular}{lccc}
&$\beta$&$\fNL$&DE\\
\hline
\rom{1}&0.05&0.0&$\Delta_{\phi} \neq 0$\\
\rom{2}&0.05&0.0&$\Delta_{\phi} = 0$\\
\rom{3}&0.0& -151.51&$\Delta_{\phi} \neq 0$\\
\rom{4}&0.0&-166.66&$\Delta_{\phi} = 0$\\
\rom{5}&0.05&-151.51&$\Delta_{\phi} \neq 0$\\
\rom{6}&0.05&-166.66&$\Delta_{\phi} = 0$\\
\hline
\end{tabular}
\caption{Different values of $\beta$ and $\fNL$ parameters used in this paper.}
\label{Table-II}
\end{table}
These values of $\fNL$ are obtained by minimizing the residual $1 - P_h/P^{\LCDM}_h$ for the combined model, i.e. they correspond to the values of maximum degeneracy for a DE-CDM coupling parameter $\beta =0.05$ for the cases $\Delta_{\phi} \neq 0$ and $\Delta_{\phi} = 0$. Clearly, the $k^{-2}$ signal, assuming $\Delta_{\phi} = 0$, is larger and therefore the amount of PNG to be degenerate with it is bigger. Therefore, for these combinations of parameters, as clearly seen in Fig. \ref{Fig-I}, IDE and PNG are strongly degenerate with each other, in the sense that their combination is {\em indistinguishable} from the fiducial $\Lambda $CDM case\footnote{We chose $\beta>0$ and $\fNL<0 $ because the same degeneracy does not apply for negative $\beta $ in the Newtonian approximation, since the coupling enters also as a $\beta^{2}$ term in Eq. \eqref{cdm_pertrub-Eq}. This means that for $\beta < 0$ and $f_{\rm NL}>0$, we do not expect a degeneracy in the non-linear regime.
}. 

Although these derived values of $f_{\rm NL}$ are at least one order of magnitude larger than currently allowed by observational constraints, we will continue to use these values as a toy example of the IDE-PNG degeneracy. Realistic values for $|f_{\rm NL}|$ with the standard assumption of a scale-independent amplitude of PNG would have too weak effects on non-linear structure formation to significantly influence the observational features for any non-negligible coupling parameter $\beta$. On the other hand, scale-dependent PNG \citep[see e.g.][]{Liguori:2010hx,Renaux-Petel:2015bja}, where $f_{\rm NL}(k)$  evolves with wavenumber $k$, may still provide an effective $f_{\rm NL}= {\mathcal O}(10^{2})$ at scales relevant for non-linear structure formation, while remaining consistent with current bounds around the Planck pivot scale $k= 0.05 h^{-1}$Mpc.

\begin{figure}
\centering
\includegraphics[width=\linewidth]{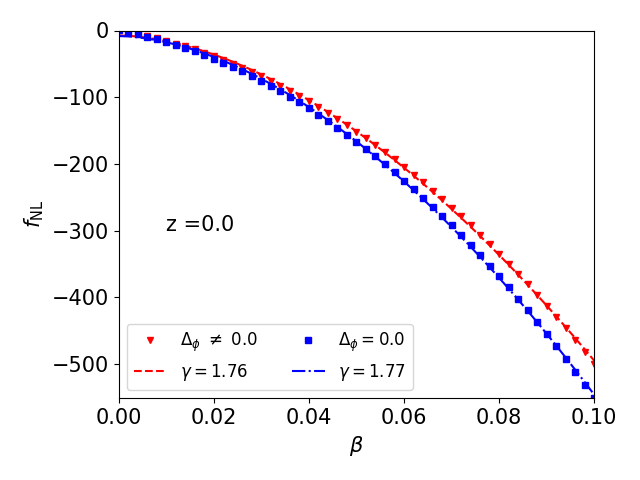}
\caption{The $\beta\text{--}\fNL$ mimicking degeneracy relation at redshift $z = 0$, over-plotted with the fitting function defined in Eq. \eqref{deg-Eq}, for the fitted values of $\gamma$.}
\label{Fig-II}
\end{figure}

In order to model the mimicking degeneracy relation between $\beta$ and $\fNL$ that is illustrated in Fig. \ref{Fig-I}, we repeat the procedure of minimizing the residual $1 - P_h/P^{\LCDM}_h$ for a wide range of the parameters $\beta$ and $\fNL$, for both perturbed and non-perturbed DE cases.  We find that  relation
\begin{ceqn}
\begin{align}\label{deg-Eq}
\fNL = \zeta\, \beta^{-\gamma},
\end{align}
\end{ceqn}
where $\zeta$ and $\gamma$ are constants, provides a good fit, with exponent $\gamma\approx 1.8$. This is shown in Fig. \ref{Fig-II}, where the numerical results are over-plotted with the fitting function Eq. \eqref{deg-Eq} for perturbed and non-perturbed DE cases. Note that the degeneracy slope $\gamma$ increases if we assume non-perturbed Dark Energy. 

\section{Degeneracy on non-linear Scales}\label{Sec-III}

It is well known that IDE and PNG separately imprint  characteristic features in the non-linear regime of structure formation, which can be tested through different observational probes. For example, IDE affects the high-mass tail of the halo mass function (HMF) by enhancing the abundance of halos \citep{Cui:2012is}, while PNG impacts the number of massive CDM halos, suppressing (increasing) it for negative (positive) $\fNL$ \citep[see e.g.][]{Grossi:2009an,Wagner:2010me}. It is therefore plausible that some form of degeneracy may appear also at these non-linear scales, and in particular that the combination of IDE with a negative value of $f_{\rm NL}$ for PNG may result in a HMF hardly distinguishable from the reference $\Lambda $CDM case at all masses.

IDE also shows distinctive features on other observational probes, including higher-order correlation functions and non-linear bias, in a similar way to PNG \citep{Moresco:2013nfa, Desjacques:2008vf, Wagner:2011wx}. IDE further affects the structural properties of CDM halos and voids \citep{Pollina:2016gsi, Pollina:2015uaa, Giocoli:2013ba, Baldi:2014tja, Baldi:2010pq}, and PNG is also expected to show significant effects on these probes \citep[][]{Neyrinck:2013iza,Abel:2011ui,Sutter:2014kda}.

This implies that the mimicking degeneracy which we have found at linear scales for the halo power spectrum may persist (fully or partly) in some small-scale non-linear observables, while it may be broken by  others. In the following, we  test the linear degeneracy relation, defined in Eq. \eqref{deg-Eq}, on non-linear scales by analysing  a suite of cosmological N-body simulations that include IDE and PNG, both separately and in a combined way. To this end, we will consider various non-linear probes, starting from the non-linear matter power spectrum and the halo-matter bias to the statistical and structural properties of CDM halos and voids. 

\subsection{N-Body Simulations}\label{Sec-NBody}

\begin{figure*}
\centering
\includegraphics[width=\linewidth]{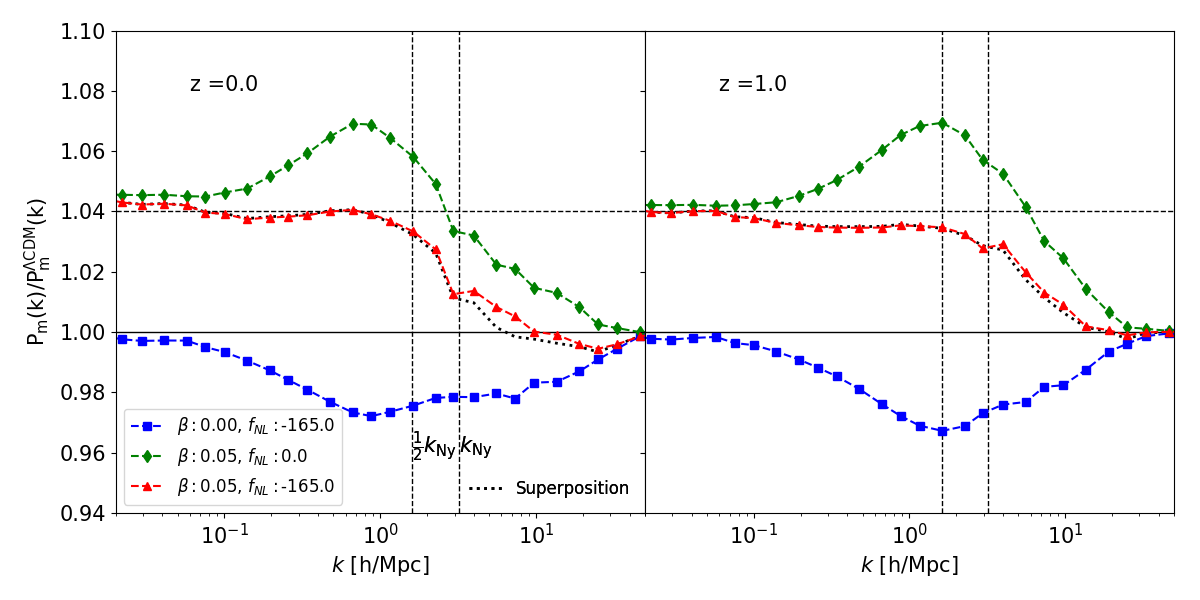}
\caption{The non-linear matter power spectrum with IDE, PNG and their combination, relative to the reference $\LCDM$ spectrum, at $z = 0$ (left panel) and $z = 1$ (right panel). The dotted black curve shows the superposition spectra (IDE-only + PNG-only). The black dashed vertical lines show the Nyqvist frequency and half of it.}
\label{Fig-V}
\end{figure*}

In order to consistently account for the effects of IDE in the non-linear regime, we made use of a modified version of the parallel TreePM N-Body code GADGET \citep{Springel:2005mi} that incorporates all the specific features of the coupling between DE and CDM, i.e. modified background expansion,  CDM particle mass time variation, the extra friction and the fifth force acting on CDM particles \citep[see][for a detailed description of the modified N-body algorithm]{Baldi:2008ay}.
The simulations follow the evolution of $1024^3$ CDM particles within a periodic cosmological box of $1\,h^{-1}$Gpc per side, for all the cosmological parameters given in Table \ref{Table-I}, with a mass resolution at $z =0$ of $5.84 \times 10^{10}\, \rm M_{\odot}/h$ and softening length $\epsilon = 24.42\, h^{-1}\mathrm{kpc}$.
Our numerical implementation of IDE assumes that DE perturbations are negligible in relation to structure formation processes compared to the dominant effects of background evolution, extra friction and fifth force. This is a valid approximation on sub-horizon scales; it becomes less accurate at scales comparable with the cosmic horizon, but this is beyond the fundamental mode of our 1$h^{-1}Gpc$ boxes. For this reason, we have chosen to consider the same approximation (i.e. $\Delta _{\phi } = 0$) to select our combination of values for the parameters $\beta $ and $f_{\rm NL}$, so as to ensure consistency between the degeneracy relation displayed in Fig.~\ref{Fig-II} and the outcomes of our N-body simulations at small scales. 

In order to generate the initial conditions for N-Body simulation of all models considered in this paper, we slightly modified the publicly available code 2LPTic \citep{Scoccimarro:2011pz}. The algorithm implements non-Gaussian initial conditions with external Hubble and growth functions consistent with IDE modifications.

The non-Gaussian initial conditions are generated for local-type PNG with an extra non-Gaussian term according to Eq. \eqref{localPNG-Eq}, where $\Phi_G$ is a random realization of a Gaussian field with the primordial power spectrum $P(k) \propto k^{n_s - 1}$. Then, the linear density field $\delta_c$ is obtained from the non-Gaussian potential $\Phi$ through the Poisson equation:
\begin{ceqn}
\begin{align}
\delta_c = \frac{2}{3} \frac{k^2}{H^2_0}\frac{D_c(z)}{\Omega_c }T(k)\Phi,
\end{align}
\end{ceqn}
where the transfer function $T(k)$ is computed using $\rm CAMB$ \citep{Lewis:1999bs} for the fiducial $\LCDM$ cosmology. We assume the transfer function is not affected  by the late-time interaction \citep{Baldi:2008ay, Baldi:2011qi}. The growth function $D_c$ for all  models is normalized at $z_{\rm CMB}\approx 1100$ to directly compare the impact of IDE on the structure growth in the period between $z_{\rm CMB}$ and the present time. For PNG, we set $\fNL = -165.0$ as the value corresponding to the interaction rate $\beta = 0.05$ on linear scales (with $\Delta _{\phi } = 0$), as determined by Eq. \eqref{deg-Eq}. 

Particle positions are then displaced from a homogeneous glass distribution \citep{Baugh:1994hb} using the Zel'dovich approximation \citep{Zeldovich:1969sb} according to the displacement field $\delta_c$ at the initial redshift $z_{i}  = 49$. In order to compute particle initial velocities, we used the relation $v(k,z) \propto f(z) \delta(k,z)$, where the growth rate function $f(z) \equiv -d\ln D_c/d\ln (1+z)$ is derived for each model by solving Eqs. \eqref{cdm_pertrub-Eq} for the growth function. For the IDE--PNG combined model, we apply the growth function of IDE after transforming the initial Gaussian potential to the non-Gaussian form according to  Eq. \eqref{localPNG-Eq}. In order to minimize the sampling variance, we used the same initial random seed for all the simulations. 
  
\section{Results}\label{Sec-IV}

In this section, we present the main results of our numerical simulations of IDE,  PNG and the combined IDE--PNG extensions rescaled with respect to the fiducial $\LCDM$ model. We focus mainly on the non-linear matter power spectrum, the halo-matter bias and the statistical and structural properties of CDM halos and voids. 

\subsection{The non-linear matter power spectrum}

We computed the non-linear matter power spectrum for each simulation by calculating the density field using a Cloud-in-Cell mass assignment on a cubic grid with the same resolution as the Particle Mesh grid used for the integration of the N-body system (i.e. $1024^3$).  According to this procedure, the non-linear matter power spectrum is determined up to the Nyquist scale,  $k_{\rm Ny} = \pi  N/L \sim 3.2  h/{\rm Mpc}$. We truncate the resulting power spectrum at the $k-$mode where the shot noise is below $20\%$ of the measured power. From the simulated power spectra, we can estimate the effects of IDE, PNG and, for the first time, the joint effects of IDE and PNG, on linear and non-linear scales at different redshifts. 

In Fig. \ref{Fig-V} we display the ratio of the non-linear matter power  spectrum for IDE, PNG and their combination, to that of the standard cosmological model, at $z =0$ and $z = 1$. The plots show the following features.

IDE with Gaussian initial conditions (dashed green curve with solid diamonds) -- shows the expected scale-dependent power enhancement at non-linear ranges due to the combined effects of the fifth force and of the extra friction associated with the DE-CDM interaction. Also, since we normalize the power spectrum at the redshift of the CMB, the normalization at linear scales ($\sigma^{\LCDM}_8 = 0.809$) is increased by about $5\%$ relative to the standard model (i.e. $\sigma^{\rm IDE}_8 = 0.825$), due to the higher linear growth rate in the IDE case \citep{Baldi:2010vv}; this is consistent with previous works \citep[see e.g.][]{Baldi:2010pq}. At higher redshift (right panel) the non-linear power spectrum enhancement due to the IDE fifth force is slightly reduced, with the peak ratio shifted towards smaller scales. 

Non-Gaussian initial conditions in $\LCDM$ (dashed blue curve with solid squares) -- shows the expected suppression of power at small scales, relative to the standard $\LCDM$ case. The deviation is larger at higher redshifts and the minimum shifts towards smaller scales, in agreement with predictions from the halo model presented in \citet{Fedeli:2009mt}. 

IDE and PNG combined (dashed red curve with solid triangles) -- the ratio no longer shows any significant scale dependence down to ranges corresponding to the location of the peak/minimum in the ratio for the two separate models. The difference in the power normalization at linear scales associated with the enhanced growth rate in IDE remains unchanged. This seems to indicate a mimicking degeneracy between IDE and PNG in the matter power spectrum on non-linear scales while there is no degeneracy on linear scales. 
Remarkably, the figure shows that there is a non-linear mimicking degeneracy for the same combination of parameters that produce mimicking degeneracy in the halo power spectrum at much larger scales, as described by  Eq. \eqref{deg-Eq} and Fig. \ref{Fig-II}.

In the figures we have also over-plotted, for comparison,  a black dotted curve representing a simple superposition of the two effects, i.e. the PNG-only (blue squares) deviation times the IDE-only (green diamonds) deviation. The very good agreement of this simple prediction with the actual power measured from the combined IDE-PNG simulation seems to indicate that the two effects  acting on structure formation are decoupled -- which suggests that full combined N-body simulations may be unnecessary in order to compute the combined power spectrum for other combinations of $\beta$ and $f_{\rm NL}$.

\subsection{Halo-Matter Bias}

\begin{figure*}
\centering
\includegraphics[width=\linewidth]{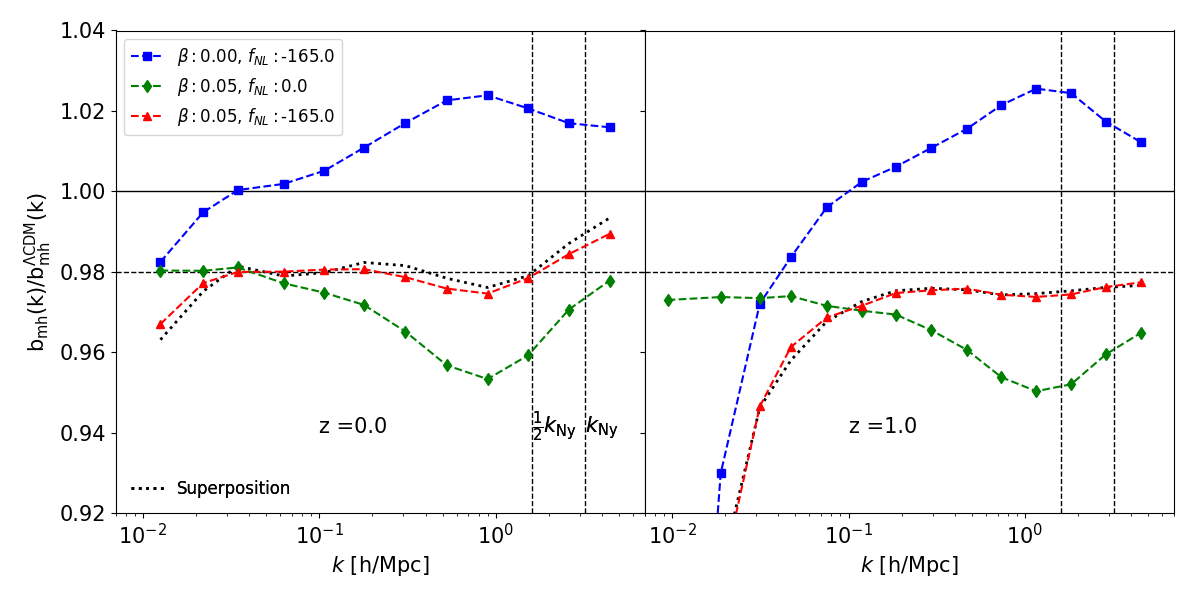}
\caption{As in Fig. \ref{Fig-V}, for the halo-matter bias. Clearly, IDE  shows no sign of scale-dependence on large scales.}
\label{FigHaloBias}
\end{figure*}

Following the standard hierarchical clustering scenario of structure formation, halos and  galaxies are biased tracers of the underlying matter distribution. In this section, we compute the linear bias between halos and the underlying dark matter density field, as the ratio between the halo-CDM cross power spectrum and the auto power spectrum in Fourier space: 
\begin{ceqn}
\begin{align}
b_{hm}(k) = \frac{P_{hm}(k)}{P_{mm}(k)}.
\end{align}
\end{ceqn}  
(We suppress the $z$-dependence for simplicity.) This bias estimator is used to avoid shot-noise \citep{Hamaus:2010im,Smith:2006ne,Baldauf:2009vj,Baldauf:2013hka}, and we follow the approach of \citet{Villaescusa-Navarro:2013pva} for the computation of the two power spectra.

In Fig. \ref{FigHaloBias}, we show the ratio of the halo-matter bias for   the IDE, PNG and IDE--PNG models, relative to the fiducial $\LCDM$ model. As expected, PNG  introduces a clear scale-dependence at large scales. On the contrary, the bias in the IDE model appears to have a slightly lower normalization than $\LCDM$ though retaining the same evolution with scale as the standard scenario. This different behaviour is most visible at higher redshifts, as shown in the right panel of Fig. \ref{FigHaloBias}, where the scale-dependence of the PNG simulation is stronger. On non-linear scales, both PNG and IDE show a maximum deviation relative to the reference model but in opposite directions, with the amplitudes of the peak/ minimum increasing and their position moving towards smaller scales at higher redshifts. 
These outcomes are all consistent with the previous literature \citep[][]{Matarrese:2008nc,Desjacques:2008vf,Moresco:2013nfa,Marulli:2011jk} and qualitatively show how the halo bias is affected at similar scales for both IDE and PNG.

For the combined IDE--PNG scenario, we find that at $z=0$ the halo bias retains some scale-dependence on large scales, i.e $k < 0.05\ h/$Mpc, while it is nearly scale-independent on scales $0.05\ h/$Mpc $<  k  < 0.5\ h/$Mpc. Furthermore, it retains the lower normalization that characterizes the IDE model at all scales. This combination of the two effects is more clear at higher redshift, where we can clearly identify two distinct regions for scale-dependent ($k<0.1\ h/$Mpc) and scale-independent ($k>0.1\ h/$Mpc) deviations from the reference model. Also in this case, the simple superposition of the two separate effects, very accurately reproduces the behaviour of the combined IDE--PNG simulation, thereby suggesting that the two phenomena act on the biasing of collapsed structures independently.

Similar to the non-linear matter power spectrum, the halo-matter bias satisfies the $\beta\text{--}\fNL$ degeneracy relation, Eq. \eqref{deg-Eq}, on non-linear scales, while this is broken at larger scales. We argue that this may be due to the fact that our N-Body implementation of IDE  as discussed above \citep[see also][]{Baldi:2010pq}, does not account for the scale-dependent growth function on large scales due to CDM--DE coupling and the contribution of large-scale DE perturbations (i.e. it assumes the approximations $D_c(k,z) \simeq D_c(z)$ and $\Delta _{\phi }=0$). 
Therefore, including the effects of large-scale CDM--DE coupling and DE perturbations should boost in a scale-dependent way the IDE linear power spectrum and consequently the halo-matter bias on large scales. This would recover the result of a mimicking degeneracy at all scales that was obtained from linear perturbation theory (Sec. \ref{Sec-II}). A proper verification of this conjecture would require major modifications to our N-body codes, that go beyond the scope of the present paper, and we defer an extensive study on this subject to future works.  

\subsection{Statistical and structural properties of CDM halos}

\begin{figure*}
\centering
\includegraphics[width=\linewidth]{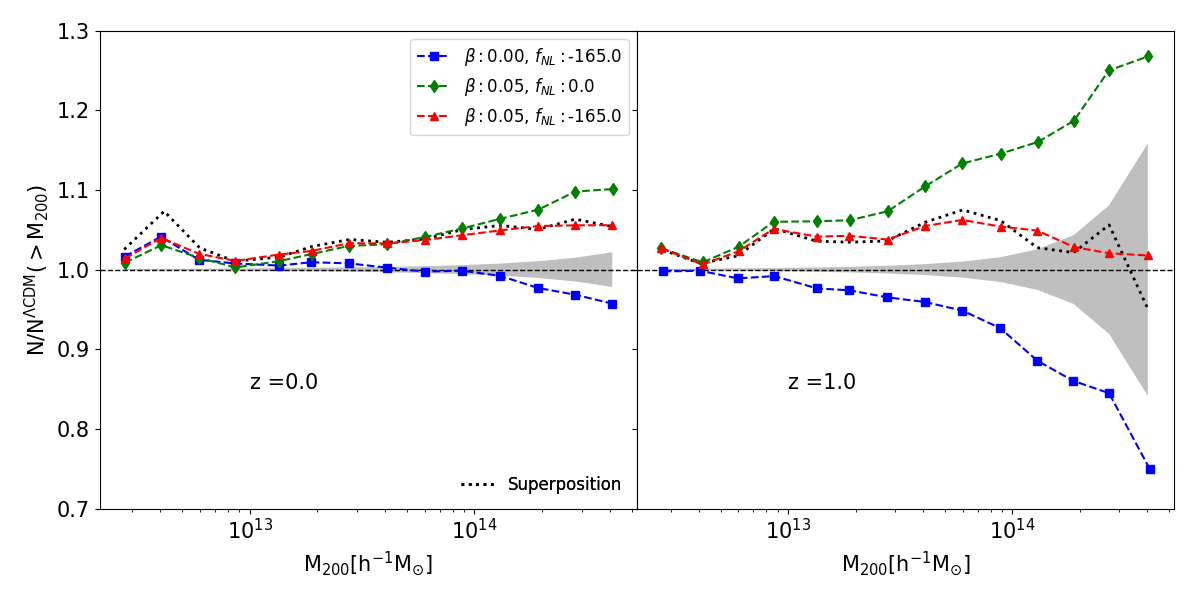}
\caption{As in Fig. \ref{Fig-V}, for the halo mass function. The grey region represents the propagated Poissonian error of the number counts of halos in each bin.} 
\label{Fig-VI}
\end{figure*}

In this section we test the $\beta\text{--}\fNL$ degeneracy relation in the statistical and structural properties of CDM  halos. 

\subsubsection{The Halo Mass Function}

We identified collapsed halos in our simulations following a standard procedure, amounting to a first identification of particle groups by means of a Friends-of-Friends (FoF) algorithm with linking length $l = 0.2 \bar{d}$, where $\bar{d}$ indicates the mean inter-particle separation. On top of these FoF halos we run the
SUBFIND algorithm \citep[][]{Springel:2000qu} in order to identify gravitationally bound sub-structures present within each group. The latter procedure allows to assign to each FoF group the virial mass $\rm M_{200}$ of its primary sub-structures, defined as the  mass of a spherical region with its centre on the particle with the halo's minimum potential enclosing a mean overdensity equal to $200$ times the critical density of the universe. 

Given these halo catalogues, we computed the halo mass function for IDE, PNG and the combined IDE--PNG models by binning the halo masses into 13  logarithmically equally-spaced mass bins over the mass range $\rm 2.0 \times 10^{12}\, M_{\odot}/h - 5.0 \times 10^{14}\, M_{\odot}/h$. The lower mass bound is set by the minimum halo mass resolved in the fiducial $\LCDM$ model, composed of at least $20$ particles.
 
In Fig. \ref{Fig-VI} we show the ratio of the cumulative HMF to the $\LCDM$ model for IDE, PNG and the combined IDE--PNG models. As expected, IDE  enhances the abundance of large mass halos with respect to the standard $\LCDM$ case, while PNG shows on the contrary  a suppression of the abundance of halos in the high-mass tail, consistent with previous results \citep{Cui:2012is, Wagner:2010me}. 

The combined IDE--PNG model shows some level of degeneracy with the standard $\LCDM$ cosmology at $z=0$, with the combined mass function being slightly lower than in the pure IDE case. The degeneracy becomes more clear, given the larger amplitude of the individual effects, at higher redshifts ($z=1$), where the IDE and PNG deviations from the reference model reach about $25-30\%$ at the largest masses with abundance suppression  by only $5\%$ in the combined case. Furthermore, the exponential dependence on halo mass of the deviation with respect to $\LCDM$ is also significantly weakened in the combined model. Nonetheless, as a mimicking degeneracy is never fully attained, the halo mass function seems not to follow the degeneracy relation of Eq. \ref{deg-Eq}, thereby providing a possible way to disentangle these phenomena. 

The simple superposition of IDE and PNG models reasonably agrees with the combined IDE--PNG simulation except in the low mass end of the halo mass function at $z = 0$ (see left panel of Fig. \ref{Fig-VI}) where some disagreement appears. This presumably could be related to the poor resolution of small mass halos thereby arising due to  numerical artefacts associated with the specific halo finder that we employed.

\subsubsection{The subhalo mass function}
\begin{figure}
\centering
\includegraphics[width=\linewidth]{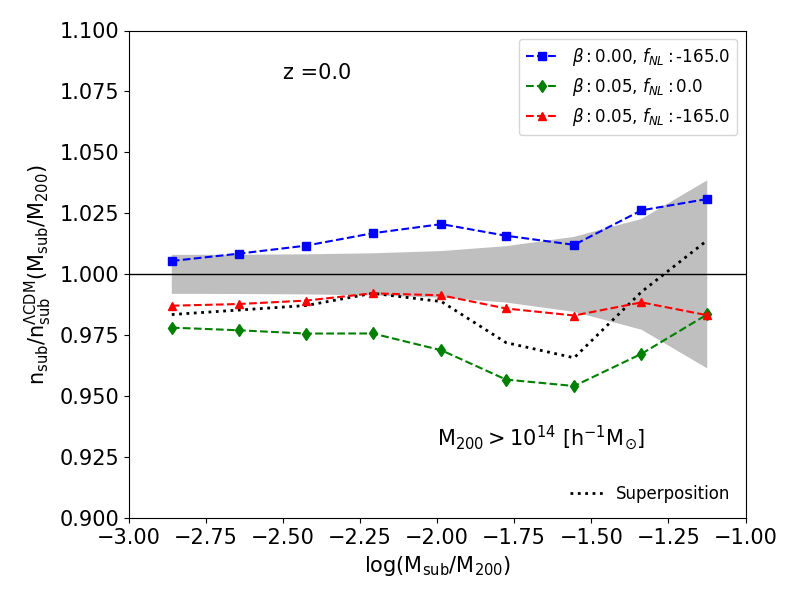}
\caption{The subhalo mass function for the cosmologies under investigation at $z= 0$. The grey region represents the propagated Poissonian error of the number counts of subhalos in each bin and the dotted black line represents the superposition of IDE and PNG models.} 
\label{Fig-VII}
\end{figure}

\begin{figure*}
\includegraphics[width=\linewidth]{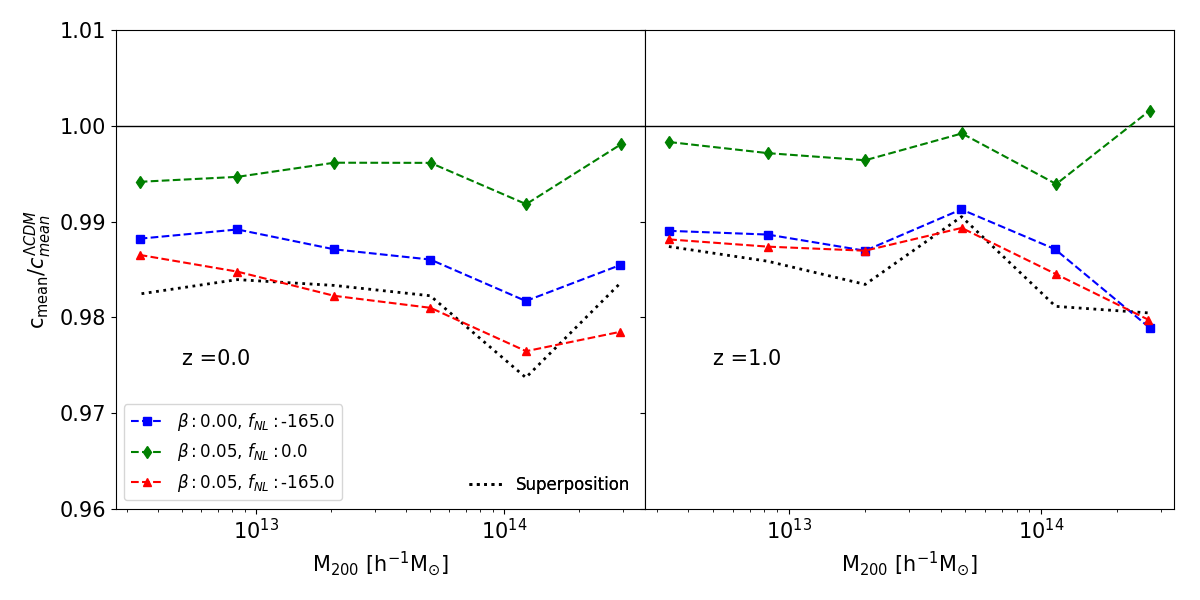}
\caption{As in Fig. \ref{Fig-V}, for the concentration-mass relation.}
\label{NFW-Fig}
\end{figure*}

As a further statistic of structure properties at small scales, we  computed -- for all our simulated cosmologies -- the subhalo mass function, defined as the number of subhalos of mass $\rm M_{\rm sub}$ within a main halo of virial mass $\rm M_{200}$. In Fig. \ref{Fig-VII}, we display the ratio of the subhalo mass function with respect to the measurements in the $\LCDM$ simulation, as a function of the mass ratio $\rm M_{\rm sub}/M_{200}$. In order to avoid resolution effects, we consider only subhalos hosted by cluster-size halos, i.e. systems with $\rm M_{200} > 10^{14} M_{\odot}/h$. We underline to the reader that the measured subhalo counts in the $\LCDM$ model are characterized by the typical slope of approximately $-1$ consistent with different previous findings \citet{gao04,giocoli10,despali17}.   

As can be seen from the figure, IDE suppresses the abundance of sub-structures over the whole range of subhalo fractional mass, even though the effect is  small (about 3-5\%). On the contrary,  PNG  enhances the abundance of subhalos up to about $4\%$ (for the highest values of the subhalo fractional mass) over the same mass range. 
The combined IDE--PNG case shows again a quite clear degeneracy, with a suppression never exceeding  $\approx 1\%$, marginally consistent with the Poissonian error range of the $\LCDM$ model. The  simple superposition of IDE and PNG models is in reasonable agreement with the combined IDE--PNG simulation. These results underline that the $\beta\text{--}\fNL$ degeneracy relation seems to remain valid also at the level of CDM halo sub-structures.

\subsubsection{Halo Concentration}

Finally, we conclude our investigation of the combined effects of IDE and PNG on structural properties of collapsed halos by computing the average halo concentration as a function of halo mass, which is usually known as the concentration-mass relation \citep{Zhao:2008wd,Giocoli:2011hz}. In order to compute the concentrations for the halos identified in our simulations, we adopt the NFW formula used in \citet{Springel:2008cc}:
\begin{ceqn}
\begin{align}\label{NFW-Eq}
\delta_{\rm con} = \frac{200}{3} \frac{c^3}{\ln (1 + c) - c/(1 + c)} = 14.426 \left( \frac{V_{\rm max}}{H_0 r_{\rm max}}\right)^2,
\end{align}
\end{ceqn}
where $\delta_{\rm con}$ is the characteristic overdensity, $c$ is the halo concentration,  $V_{\rm max}$ is the maximum circular velocity of the halo attained at radius $r_{\rm max}$.  In Fig. \ref{NFW-Fig}, the ratio of the concentration-mass relation of IDE, PNG and the combined IDE--PNG models relative  to $\LCDM$ is presented at $z = 0$ (left panel) and $z = 1$ (right panel). 

As expected, IDE halos are found to be less concentrated with respect to the fiducial $\LCDM$ case, in agreement with results given in \citep{Baldi:2010pq}. Similarly, PNG with  $f_{\rm NL}<0$ also suppresses halo concentrations (the opposite would occur for a positive $f_{\rm NL}$). Therefore, for the first time we encounter an observational probe showing deviations from $\Lambda $CDM pointing in the same direction for IDE and our negative $f_{\rm NL}$ PNG scenarios.

The combined IDE--PNG simulation, accordingly, shows an even stronger suppression of the concentration-mass relation relative to the $\LCDM$ model than the two individual models separately. The effects are less pronounced at higher redshifts, while the trends and the relative ordering of the various models is preserved. Superposition of the individual effects of IDE and PNG seems to agree well with the combined simulation.  This however indicates that the  $\beta\text{--}\fNL$ degeneracy is broken for the CDM halo concentration-mass relation, which might then represent another direct way to disentangle the models, when combined with another more degenerate probe. It is also reasonable to emphasize that this effect is relatively small; only future wide field observational campaigns -- like the future ESA-mission Euclid \citep{Laureijs:2011gra} -- will be able to collect the large number of galaxy groups and clusters \citep{Sartoris:2015aga} necessary for these tests. 

\subsection{Statistical and structural properties of cosmic voids}

\begin{figure}
\includegraphics[width=\linewidth]{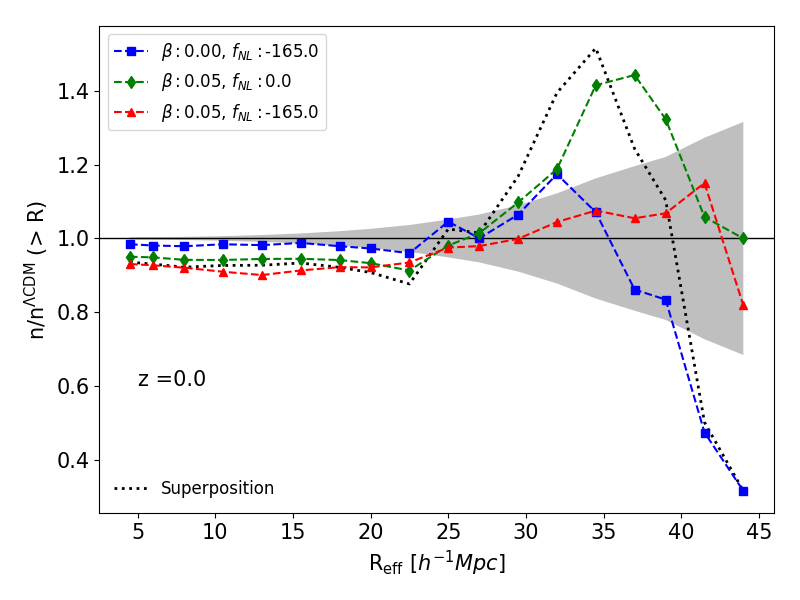}
\caption{As in Fig. \ref{Fig-VII}, for the void number function.}
\label{Fig-VNF}
\end{figure}

\begin{figure*}
\includegraphics[width=\linewidth]{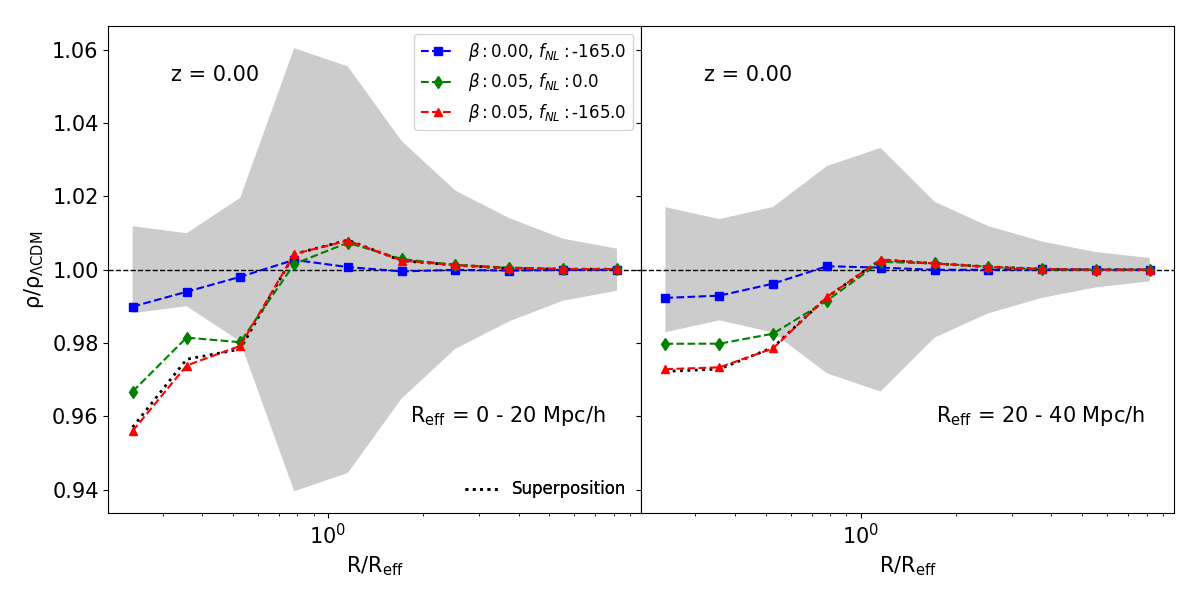}
\caption{The ratio of the stacked void density profiles in two different ranges of effective radius $R_{\rm eff} = 0 - 20$ (left panel) and $ R_{\rm eff} = 20 - 40$ (right panel) to the standard model at $z = 0$. The grey region represents the $2\sigma$ bootstrap standard deviation of 1000 re-sampled profiles.}
\label{Fig-VDP}
\end{figure*}

In this section, we  move our focus to under-dense regions of the universe by testing whether cosmic voids also follow the $\beta\text{--}\fNL$ degeneracy relation. In order to identify cosmic voids in our set of simulations, we employ the publicly available void finder VIDE \citep{Sutter:2014haa}, which is based on the ZOBOV algorithm \citep{Neyrinck:2007gy}. The cosmic void identification is mainly done by means of a Voronoi tessellation scheme that associates a polyhedrical cell to each particle tracing the CDM density field. Subsequently,  cell volumes are compared in order to identify local density minima, i.e. cells with a larger Voronoi volume than all their surrounding cells. A hierarchy of identified voids is then obtained via the watershed transform algorithm \citep{Platen:2007qk}, by joining Voronoi cells around a local density minimum.  In our analysis, we consider only voids with a central density that is below the density of the universe by $20 \%$ and a lower density contrast limit $1.57$, corresponding to a probability of voids arising from Poisson noise below $\sim 5\%$, i.e only voids at $2\sigma$ confidence level are considered  \citep{Neyrinck:2007gy}.  

\subsubsection{Void number function}

As a first statistics for cosmic voids, we study their abundance as a function of the void effective radius $ R_{\rm eff}$, defined as the radius of a sphere centred on the most underdense particle of a void and having the same volume as the Voronoi volume of the void:
\begin{ceqn}
\begin{align}
V_{\rm void} \equiv \sum_{i=1}^{N} V^p_i= \frac{4}{3} \pi R^3_{\rm eff}.
\end{align}
\end{ceqn} 

In Fig. \ref{Fig-VNF} we show the ratio of the void number functions relative to the ones in the $\LCDM$ cosmology for all models under consideration, as a function of the effective radius $R_{\rm eff}$ at $z = 0$. From the figure we see that IDE suppresses the number of cosmic voids with effective radius $R_{\rm eff} < 25 h^{-1}\,$Mpc by about $5\%$ relative to the $\LCDM$ case, and correspondingly  enhances the abundance of larger voids by up to $40\%$. The trend is qualitatively similar, though quantitatively weaker (up to ranges $R_{\rm eff}\approx 30~\rm Mpc/h$), for PNG, in agreement with previous results of \citet{Kamionkowski:2008sr}. However PNG strongly suppresses void number function at $R_{\rm eff} \approx 40~\rm  Mpc/h$, while IDE enhances it by $40\%$ at the same scales. 

The combined IDE--PNG simulation shows suppression of the void number function for radii $R_{\rm eff} < 25h^{-1}\,$Mpc, similar to the IDE case and barely enhances the void abundance at $R_{\rm eff} > 25h^{-1}\,$Mpc relative to the $\LCDM$ case, so that it is indistinguishable within the $\LCDM$ Poisson error range at these radii. As we did for all previous observables, we also compute the simple superposition of the two effects, by taking the product of the two separate deviations with respect to the reference case. For the first time in our analysis, we  see that such a superposition fails to reproduce the results of the combined simulation at large void effective radii: this follows from comparing the black dotted curve, representing the analytical superposition, with the blue squares, showing the combined simulation in Fig. \ref{Fig-VNF}. In this case we notice that the simple superposition of the two fields tends to be mainly dominated by the IDE not leaving much contribution to the PNG. This suggest that in the full simulation, in void regions, a cross-talk term between the two non-standard extensions emerges moving down the void number counts with respect to the simple superposition.

This suggests that the two phenomena interplay in some way in shaping the growth of large cosmic voids, and cannot be considered as fully independent in this regime.
In any case, we notice that the $\beta\text{--}\fNL$ degeneracy is fulfilled by the abundance of cosmic voids with large effective radii ($R_{\rm eff} > 25h^{-1}\,$Mpc), while it does not seem to apply at smaller void radii. 

\subsubsection{Void density profiles}

To further check the $\beta\text{--}\fNL$ degeneracy on cosmic void structural properties, we computed the average void density profiles for two different bins of void radius, namely $0  < R_{\rm eff} < 20h^{-1}\,$Mpc and $20 < R_{\rm eff} < 40h^{-1}\,$Mpc. We do this by stacking individual density profiles of $100$ randomly selected voids, for each radius bin, corresponding among the different cosmological simulations.  We display the ratio of the resulting void mean density profiles in Fig. \ref{Fig-VDP} for all considered models, relative to $\LCDM$ at $z = 0$. The grey area represents the $2\sigma$ confidence limit, computed by means of a {\em bootstrap} re-sampling technique.

Again, we compare the observational signature of the individual IDE and PNG models with their combination. As can be seen from the plot, cosmic voids in the IDE case tend to have a lower inner density than their $\LCDM$ counterparts.  This indicates that cosmic voids are emptier in the IDE case, fully consistent with previous results \citep[see e.g.][]{Pollina:2015uaa}. Correspondingly, the  compensating over-density around the effective radius $R_{\rm eff}$ is found to be more prominent than in $\LCDM$. On the other hand, PNG shows a negligible effect on cosmic void density profiles. It is then not surprising that the combined IDE--PNG model also shows lower density profiles in the central regions of the voids. This result also shows  that cosmic voids do not seem to follow the same degeneracy relation that applies for most of the observables related to properties of the over-dense regions of the universe.

\section{Discussion and Conclusions}\label{Sec-V}
 
The concept of observational degeneracy in cosmology arises in several different forms: (1) \emph{Parameter Degeneracy} represents the existence of large error correlations between different model parameters for specific measurements \citep{Efstathiou:1998xx,Crooks:2003pa,Tereno:2004xe,Howlett:2012mh}; (2) \emph{Dark Degeneracy} reflects the fact that gravitational experiments measure the energy-momentum tensor of the total dark sector and splitting into Dark Energy and Dark Matter is arbitrary \citep{Kunz:2007rk, Aviles:2011ak}; (3) \emph{Mimicking Degeneracy} occurs when cosmological models different from the standard $\LCDM$ mimic some of its specific features, like background expansion and the growth of matter perturbations \citep{Fay:2007uy, Setare:2013xh, Fay:2016yow}.

Cosmic degeneracy of IDE has been investigated in the literature \citep{Clemson:2011an, Valiviita:2015dfa}, including the partial mimicking degeneracy of IDE and MG \citep{Wei:2008vw,Koyama:2009gd, Wei:2013rea}.  
A mimicking degeneracy between PNG in the power spectrum in the Newtionian approximation, and the correct general relativistic power spectrum with Gaussian initial conditions, has been shown by \citet{Bruni:2011ta,Jeong:2011as}. Also, parameter degeneracy has been investigated in the non-Gaussian halo bias by \citet{Carbone:2010sb}. 
Moreover,  \cite{Abramo:2017xnp} investigated the degeneracy of large-scale velocity effects on galaxy clustering with the (local) non-Gaussianity parameter  $f_{\rm NL}$,  by simulating galaxy surveys and combining the clustering of different types of tracers of large-scale structure. They studied how large-scale velocity contributions could be mistaken for the signatures of primordial non-Gaussianity \citep[see also][]{Raccanelli:2013dza,Raccanelli:2016avd}. 

In this paper -- as part of a Cosmic Degeneracies paper series \citep{Baldi:2013iza,Baldi:2016oce} -- we have considered the mimicking degeneracy between IDE and PNG that was first shown in linear perturbation theory by \cite{Hashim:2014rda}. Since IDE can mimic PNG, the possibility exists that we can choose IDE and PNG parameters such that the two effects cancel, i.e., produce standard $\LCDM$ behaviour.  We confirmed this mimicking degeneracy in the halo power spectrum on very large scales, i.e. $k \lesssim k_{\rm eq}$, based on purely analytical calculations in the linear regime. We then fitted the degeneracy relation with a power law, $\fNL\propto \beta^{-\gamma}$ (depicted in Fig. \ref{Fig-II}), by minimizing the residual of the halo power spectrum for the combined IDE--PNG model with respect to the mimicked $\LCDM$ model. 

To further investigate and validate the $\beta\text{--}\fNL$ degeneracy, Eq. \eqref{deg-Eq}, at non-linear scales, we employed a suite of specifically designed N-Body simulations including the effects of IDE and PNG, both separately and combined with each other. In order to increase the effects under investigation and more easily detect their signatures we chose very large  values of the PNG parameter $f_{\rm NL}$, which are already ruled out by the most recent CMB observations. Still, such values could be achieved at the scales tested by our simulations for simple extensions of the PNG model such as e.g. a scale-dependent $f_{\rm NL}$. We extracted from our simulations a set of standard statistics, and we studied their deviations from the reference Gaussian $\Lambda $CDM model. In particular, we did investigate: 
\begin{itemize}
\item[--] \emph{The non-linear matter power spectrum}, for which we observed that the mimicking degeneracy persists, remarkably, on non-linear scales in the sense that the scale-dependent deviation with respect the reference $\LCDM$ scenario characterising the two separate models at non-linear scales disappears in the combined simulation even though the difference in the linear power normalisation due to the enhanced growth rate in IDE is not removed;
\item[--] \emph{The halo matter bias}, for which we find similarly to the non-linear power spectrum, that the scale-dependence imprinted by the two different models at non-linear scales is also strongly suppressed in the combined simulation while on linear scales such scale-dependent feature is retained and so breaks the observed degeneracy;
\item[--] \emph{The halo mass function}, which also shows some level of degeneracy though not satisfying Eq. \eqref{deg-Eq} for the degenerate $\beta\text{--}\fNL$ values thus allowing us to disentangle the observed degeneracy;
\item[--] \emph{The subhalo mass function,} also showing mimicking degeneracy over the whole subhalo mass range availabe in our simulations; 
\item[--] \emph{The halo concentration-mass relation,} which we found to be the first observable to explicitly break the degeneracy as both PNG and IDE have qualitatively the same impact on halo concentrations, namely to suppress concentrations at a given mass with respect to the reference $\Lambda $CDM scenario; 
\item[--] \emph{The void number function} showing mimicking degeneracy for large voids ($R_{\rm eff} > 25-30 h^{-1}$Mpc) while the degeneracy is broken for smaller void radii; 
\item[--] and \emph{The void density profiles} for which, similarly to the case of the concentration-mass relation, the mimicking degeneracy is also not observed at all as both individual models predict a lower inner density of cosmic voids compared to $\Lambda $CDM.
\end{itemize}

Therefore, we conclude that measurements of CDM halo and cosmic void internal structural properties, namely  halo concentration-mass relation and void density profile would allow us basically to break the degeneracy when combined to any of the other probes that we investigated in this work. 

In principle, this degeneracy creates difficulties in identifying the simultaneous presence of IDE and PNG, and in accurately constraining them separately. However, in practice, the degeneracy only arises for values of $|\fNL|$ that are ruled out by current constraints. Nevertheless, our investigation has shown which non-linear probes could be most useful for improving constraints on IDE and PNG. 

\section*{Acknowledgements}
MH, CG and MB acknowledge support from the Italian Ministry for Education, University and Research (MIUR) through the SIR individual grant SIMCODE, project number RBSI14P4IH. 
CG acknowledges support from the Italian Ministry of Foreign Affairs and International Cooperation, Directorate General for Country Promotion.
The simulations described in this work were done on the Sciama High Performance Compute (HPC) cluster which is supported by 
the ICG, SEPNet and the University of Portsmouth.. 
DB acknowledges partial financial support by ASI Grant No. 2016-24-H.0. During the preparation of this work DB was also supported by the Deutsche Forschungsgemeinschaft through the Transregio 33, The Dark Universe and Unidad de Excelencia ``Mar\'ia de Maeztu".
RM is supported by the South African SKA Project and the National Research Foundation of South Africa (Grant No. 75415), and by the UK Science \& Technology Facilities Council (Grant No. ST/N000668/1).



\bibliographystyle{mnras}
\bibliography{CosmicDegen} 



%
%
%

\bsp	
\label{lastpage}
\end{document}